# Anti-reflection coatings for highly anisotropic materials in the mid infrared

HONGYAN MEI,[1,†] JIN-WOO CHO,[1,†] JAE-SEON YU[2], HUANDONG CHEN[3], SHANTANU SINGH[3,4], BOYANG ZHAO[3], JAYAKANTH RAVICHANDRAN[3,4], SUN-KYUNG KIM[2], MIKHAIL A. KATS[1,*]

[1] *Department of Electrical and Computer Engineering, University of Wisconsin-Madison, Madison, WI 53706, USA*

[2] *Department of Applied Physics, Kyung Hee University, Gyeonggi-do 17104, Republic of Korea*

[3] *Mork Family Department of Chemical Engineering and Materials Science, University of Southern California, Los Angeles, CA 90089, USA*

[4] *Core Center for Excellence in Nano Imaging, University of Southern California, Los Angeles, CA 90089, USA*

[†] *These authors contributed equally.*

[*] *mkats@wisc.edu*

**Abstract:** We develop and optimize thin anti-reflection coatings (ARCs) for highly anisotropic materials in the mid infrared. Unlike conventional ARCs that assume nearly isotropic refractive indices, this work fully integrates the anisotropic nature of materials into the design process. We describe two designs of thin ARCs for highly anisotropic materials: a single form-birefringent layer, and a planar bilayer. We realized the planar bilayer ARC experimentally, demonstrating excellent mid-infrared anti-reflectance across over a broad range of angles for all polarizations.

## 1. Introduction

Anti-reflection coatings (ARCs) reduce surface reflections and increase light transmission across interfaces in optical systems. While the simplest ARCs for conventional materials and near-normal incidence comprise a single transparent quarter-wave layer with a refractive index that is the geometric mean of the substrate and superstrate[1], most high-performance ARCs at visible and near-infrared wavelengths use multiple (e.g., 3–15) layers to achieve higher performance or additional functionalities[2–6].

The multilayer approaches can be used for wavelength ranges beyond the visible and near infrared, but with some practical limitations; for example, in the mid-infrared (mid-IR), there are fewer highly transparent materials that can be readily deposited compared to shorter wavelengths, and each layer must be an order of magnitude thicker due to the longer wavelength. Example mid-IR ARCs that have been recently demonstrated include a four-layer ARC for silicon (Si) using zinc sulfide (ZnS), yttrium fluoride ($YF_3$), and germanium (Ge), totaling 2.5 micron and achieving < 1% reflection over the range of 7–12 μm[7], and a 11-layer coating using $YF_3$ and ZnS, totaling ~2 μm, and achieving broadband anti-reflectance for 1.5–15 μm[8]. In addition to their relevance for transmissive optical components[7,8], the availability of mid-IR ARCs is also useful for applications requiring high thermal emissivity[9,10].

Our paper tackles the specific challenge of ARCs for highly anisotropic materials in the mid-IR, which comes up when trying to incorporate optical components comprising these anisotropic materials into optical systems. For example, it may be desirable to create an ultrathin mid-IR waveplate from a highly birefringent material, such as one of the $A_{1+x}BX_3$ quasi-1D chalcogenides, $BaTiS_3$[11-14] and $Sr_{9/8}TiS_3$[15,16], but this can only be done if reflection for all polarizations is suppressed at the same time.

In $BaTiS_3$, the difference in refractive index ($\Delta n = n_e - n_o$) along different directions can be as large as $\Delta n > 0.75$, and in $Sr_{9/8}TiS_3$ $\Delta n$ can exceed 2 [Fig. 1]. These materials also possess high refractive indices, making ARCs essential for their use in any practical optical system. These large, direction-dependent refractive index values result in varying Fresnel reflectance and transmittance based on light polarization and propagation direction, complicating ARC design. In an isotropic material, a single ARC optimized for one refractive index works for all polarizations (at least close to normal incidence). As $\Delta n$ increases, however, the two axes present increasingly different optical interfaces, and a coating optimized for one axis leaves significant residual reflection along the other optical axis. This challenge is further complicated in the mid-IR, where the pool of transparent materials is more limited[17], and compounded by the longer wavelengths that require thicker ARC layers compared to visible-range ARC designs.

Here, we designed two ARC structures to minimize the number of layers while achieving good anti-reflection performance: a single-layer structure that requires lithographic patterning, and a planar bilayer structure that can be vapor-deposited with no lithography. We fabricated the bilayer structure as an ARC for $BaTiS_3$, which demonstrated robust performance over broad infrared spectral ranges and wide incident angles for both ordinary and extraordinary polarization. To the best of our knowledge, this is the first demonstration of an ARC designed to address the unique challenge of highly anisotropic materials. Our findings therefore offer design guidelines for ARCs tailored for anisotropic materials.

## 2. Theory and calculations

The quasi-one-dimensional materials $BaTiS_3$ and $Sr_{9/8}TiS_3$ have some of the largest optical birefringence measured in the mid-IR [Fig. 1], resulting from a combination of anisotropic $ABX_3$ structure and picometer-scale atomic displacements in the case of $BaTiS_3$[11,12], and structural modulations resulting in enhanced anisotropy in the polarizability in the case of $Sr_{8/9}TiS_3$[15]. In these anisotropic materials, the interface (Fresnel) reflectance has large polarization dependence even at normal incidence [Figs. 1(c, f)].

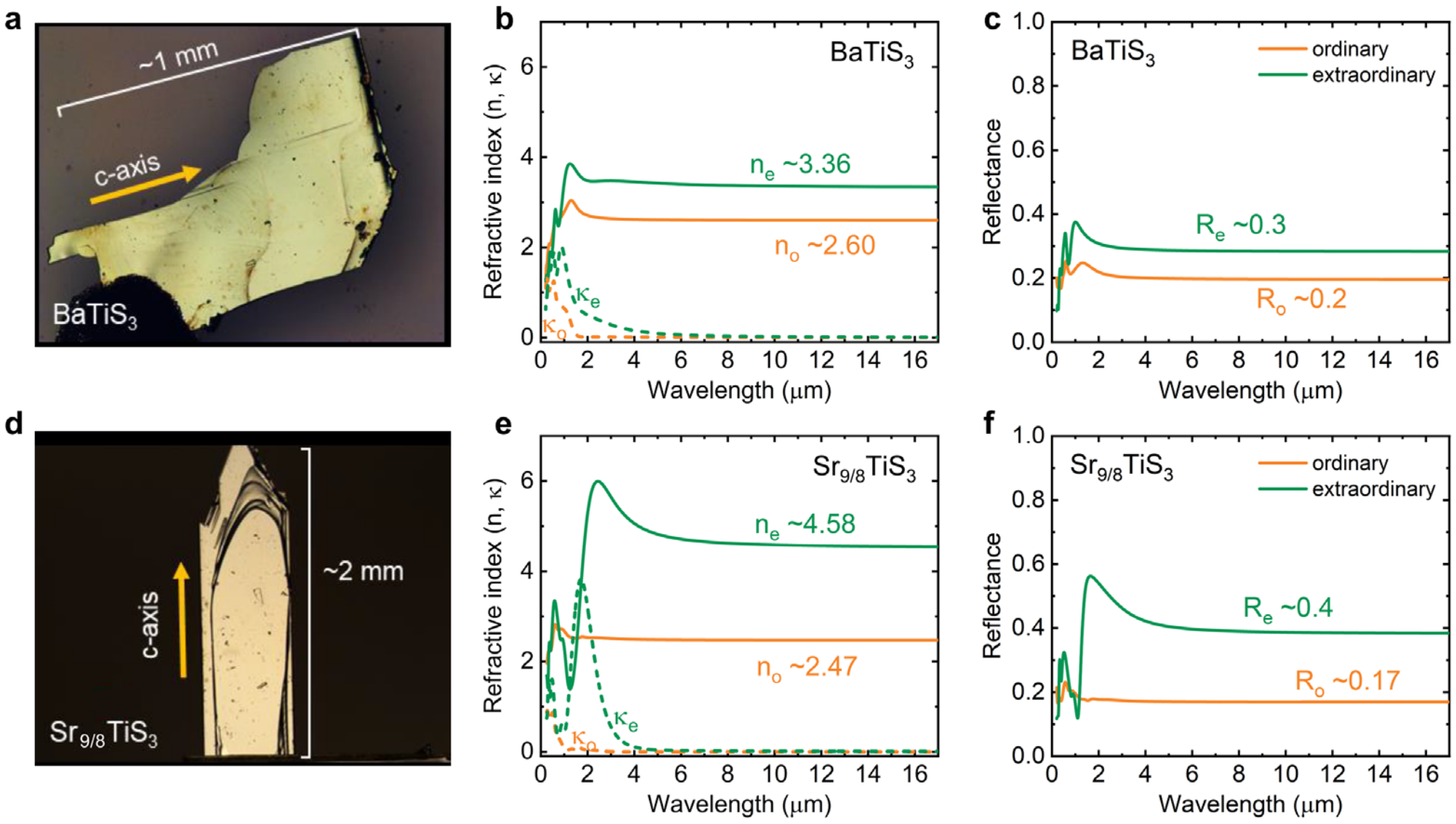

**Figure 1.** (a, d) Optical images of (a) a $BaTiS_3$ crystal, and (d) a $Sr_{9/8}TiS_3$ crystal. The *c*-axis lies in plane along the direction indicated by the orange arrow. (b, e) Complex refractive indices for (b) $BaTiS_3$ and (e) $Sr_{9/8}TiS_3$, from 210 nm to 17 μm, for light polarized parallel (extraordinary

direction, $n_e$) and perpendicular (ordinary direction, $n_o$) to the $c$-axis. The $BaTiS_3$ panels are adapted from ref.[11], and the $Sr_{9/8}TiS_3$ panels are adapted from ref.[15]. (c, f) Calculated normal-incidence reflectance at an interface between air and (c) $BaTiS_3$ and (f) $Sr_{9/8}TiS_3$.

### *2.1 Single-layer form-birefringent ARC*

The simplest conventional ARC to air is composed of a single thin layer of transparent material, whose refractive index is the square root of the substrate's refractive index, with a thickness equivalent to a quarter of a designated wavelength. For birefringent substrates like $BaTiS_3$ and $Sr_{9/8}TiS_3$, the single layer must meet the "quarter-wave" condition for both the ordinary and extraordinary axes simultaneously, while maintaining a uniform thickness, which cannot be achieved with isotropic films. Therefore, we designed a form-birefringent[18] anisotropic ARC, comprising subwavelength ridges [Figs. 2(a, d)].

Effective medium theory (EMT) can be used to aid with the design of such form-birefringent structures. For subwavelength ridges with a rectangular cross-section, the parallel and perpendicular effective refractive index can be approximated as[19,20]:

$$\tilde{n}_{\parallel} = [(1-D)\tilde{n}_1^2 + D\tilde{n}_2^2]^{1/2} \tag{1}$$

$$\tilde{n}_{\perp} = \frac{\tilde{n}_1 \cdot \tilde{n}_2}{[D\tilde{n}_1^2 + (1-D)\tilde{n}_2^2]^{1/2}} \tag{2}$$

where $D$ is the duty cycle, $n_1$ is the index of the matrix (e.g., air), and $n_2$ is the index of the material comprising the ridges. In our design, we position the ridges in the direction of the $c$-axis of the substrate $ABX_3$ material [Figs. 2(a, d)].

We first calculate the geometric mean of the indices of the substrate ($ABX_3$ material) and superstrate (air) along two different orientations, yielding the required $n_{||}$ and $n_{\perp}$ for the ARC. We then plug these into the EMT equations (Eqn. 1 and Eqn. 2), along with $n_1 = 1$ (air), and solve the two equations simultaneously to get the index of the ideal ridge material, $n_2$, and the corresponding duty cycle $D$. To meet the quarter-wave ARC condition for both polarizations at normal incidence, we calculate the thickness $t$ by using the average index $(n_1 + n_2)/2$, targeting a wavelength of 10 μm. We then selected a period for the ridges of 2 μm, which is subwavelength but large enough for conventional photolithography. This calculation for $BaTiS_3$ at $\lambda = 10$ μm, yields $n_2 \approx 1.96$, $D \approx 0.8$, $t \approx 1.60$ μm, whereas for $Sr_{9/8}TiS_3$, $n_2 \approx 2.45$, $D \approx 0.71$, $t \approx 1.35$ μm.

We carefully reviewed possible transparent materials for vapor deposition and identified each material's transparency window, here defined by an absorption coefficient of $< 10$ $cm^{-1}$, for material selection in the design, as shown in Fig. 3. Based on the optimum $n_2$ values obtained from the EMT calculation, ZnS ($n \sim 2.2$) and ZnSe ($n \sim 2.4$) were chosen to be the most suitable materials to achieve the target $n_2$ values for $BaTiS_3$ and $Sr_{9/8}TiS_3$, respectively. They exhibit high refractive indices and broad transparency windows across the wavelength range of interest (highlighted area in the Fig. 3a).

To fine-tune our design, we performed full simulations using the finite-difference time-domain (FDTD) method to obtain precise values of the duty cycle and thickness. For $BaTiS_3$, we found a combination of air and ZnS with a duty cycle of $D = 0.725$ and a thickness of $t = 1.35$ μm [Figs. 2(a–c)]. For $Sr_{9/8}TiS_3$, we identified a combination of air and ZnSe with a duty cycle of $D = 0.7$ and a thickness of $t = 1.25$ μm [Figs. 2(d–f)].

The resulting ARCs suppress reflectance at $\lambda = 10$ μm to $R < 0.01$, though notably the reflectance minimum is at a slightly different wavelength for the two polarizations [Fig. 3(c, f)]. If targeting a particular wavelength, this is an effective ARC strategy. As an example, we found that by applying this single-layer ridge structure to both the top and bottom surfaces of a finite-thick (10 μm) $BaTiS_3$ and $Sr_{9/8}TiS_3$ crystals, interference-induced Fabry–Pérot fringes can be substantially mitigated at the design wavelength, as shown in Figs. S1 and S2 in Supplementary Information.

We only show the simulation results for the single-layer form-birefringent ARCs because the experiments were challenging given the sizes and irregular shapes of the $BaTiS_3$ and $Sr_{9/8}TiS_3$ samples we were able to grow at the time of this writing. Our samples are hundreds of microns in size [Figs. 1 and 5], making it difficult to uniformly spin photoresist, resulting in further challenges in liftoff of relatively thick ZnS or ZnSe to form the necessary ridges. Nevertheless, we believe that the form-birefringent ARCs can be readily fabricated on larger crystals.

We note that anisotropic ARCs similar to ours have previously been explored in a different context, to suppress reflectance for one polarization but not for another one, assuming an isotropic substrate[21].

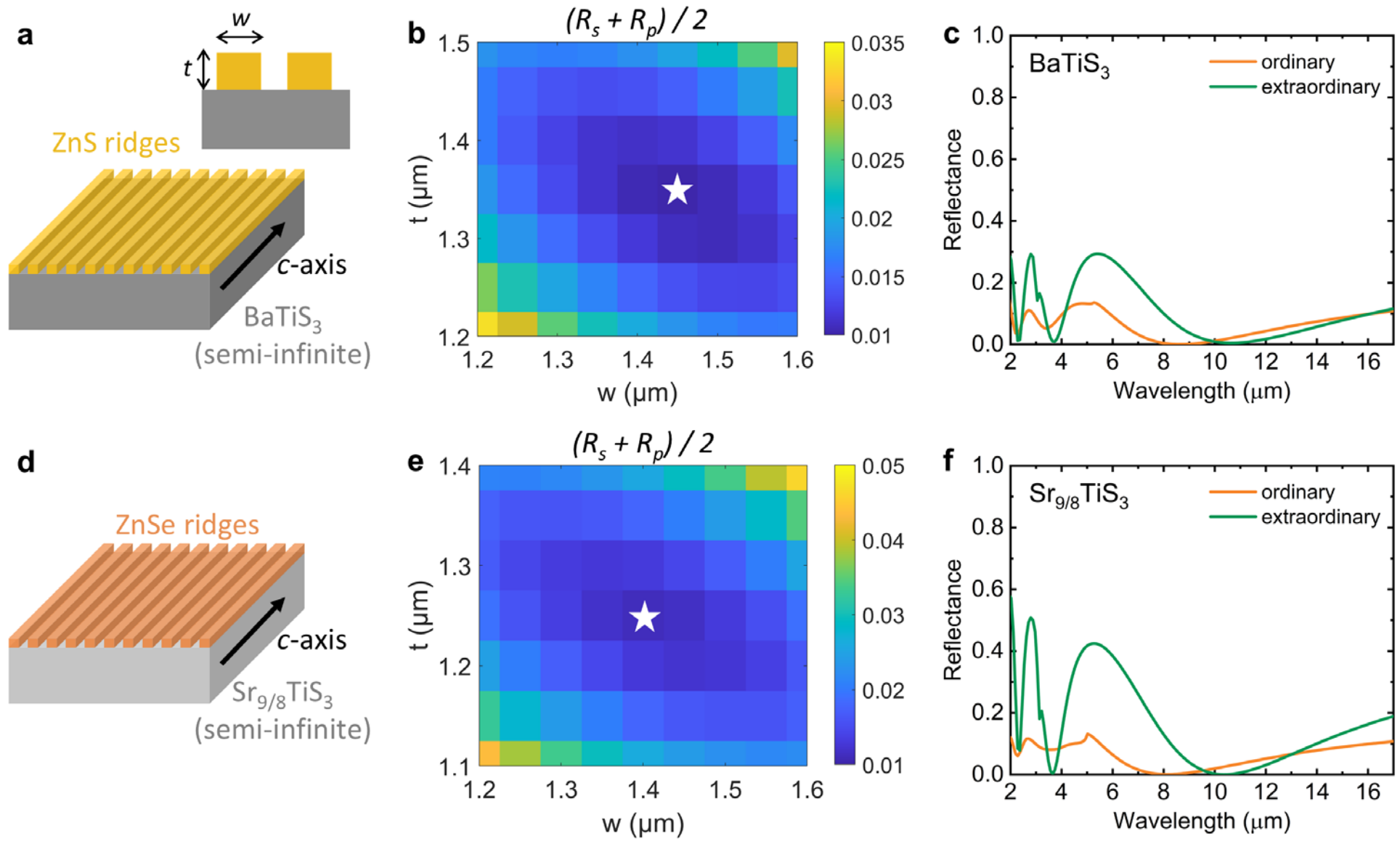

**Figure 2.** Form-birefringent single-layer ARC using ridges comprising an isotropic material. (a) ZnS ridges on a $BaTiS_3$ substrate, with ridges oriented along the *c*-axis. (b) Calculated polarization-averaged reflectance, $(R_s + R_p)/2$, at λ = 10 μm as a function of ridge thickness ($t$) and width ($w$). The minimum average reflectance occurs at $w$ = 1.45 μm and $t$ = 1.35 μm, with a period for the ridges of 2 μm. (c) Calculated $R_s$ and $R_p$ spectra for the corresponding $w$ and $t$ parameters. (d) ZnSe ridges on an $Sr_{9/8}TiS_3$ substrate, with ridges oriented along the *c*-axis. (e) Calculated polarization-averaged reflectance, $(R_s + R_p)/2$, at λ = 10 μm as a function of $t$ and $w$. The minimum average reflection occurs at $w$ =1.4 μm and $t$ = 1.25 μm. (f) Calculated $R_s$ and $R_p$ spectra for the corresponding $w$ and $t$ parameters.

### *2.2 Planar bilayer ARC*

To design a single-layer ARC for an isotropic window, two degrees of freedom (i.e., thickness and refractive index) are sufficient to suppress the reflectance at normal incidence. For an anisotropic window or other substrate, additional degrees of freedom are needed. In Section 2.1, we found that adding a third degree of freedom, the duty cycle of ridges, over an isotropic single layer gives different effective indices for the two optical axes and suppresses the reflectance for both polarizations. However, the reflectance minima for the ordinary and extraordinary axes occur at different wavelengths, λ = ~8 μm for ordinary axis and λ = ~11 μm for the extraordinary axis [Fig. 2], and broadband suppression for both polarizations cannot be achieved simultaneously without adding yet another degree of freedom. This limitation, together with the need for lithographic patterning, motivates a planar bilayer ARC design that requires no patterning while providing the additional degrees of freedom needed to achieve broadband reflection suppression for both axes.

To increase the number of degrees of freedom available in a planar ARC, we can introduce an arbitrary number of additional layers. However, we also want to keep the total thickness small, since mid-IR wavelengths are substantial compared to typical layer thicknesses that can be easily deposited. This leads to a trade-off between the number/total thickness of layers and the effectiveness of the ARC.

We focused on a bilayer structure to demonstrate a design rule for ARCs for anisotropic materials. Figure 3 shows various low- and high-index materials that are transparent in the mid-IR and can be vapor-deposited. For a target wavelength range from 6 to 10 μm, fluorides ($CaF_2$, $YF_3$, $BaF_2$) feature a low refractive index with a broad transparency window, and ZnS, ZnSe, $As_2S_3$, CdTe, amorphous Si (a-Si), and amorphous Ge (a-Ge) are suitable high-index materials with low loss. Out of these, we selected $YF_3$ as the low-index material and ZnS, ZnSe, a-Si, and a-Ge as the high-index materials, because we had access to a physical vapor deposition system that accommodated all of these materials.

We selected $BaTiS_3$ as the highly anisotropic substrate material, and considered a bilayer ARC on the $BaTiS_3$ surface. For each possible combination of $YF_3$ with a high-*n* material (ZnS, ZnSe, a-Si, and a-Ge), we swept through all possible thicknesses of each layer to minimize the normal-incidence reflectance, averaged over the two polarizations, targeting $\lambda = 6$–$10$ μm. The best design for each combination is shown in Fig. S3. Among these, a combination of a 1.41-μm-thick $YF_3$ layer and a 0.85-μm-thick ZnS layer can reduce the reflectance of $BaTiS_3$ to < 2% for both polarizations in the range of 6–10 μm, as shown in Fig. 4.

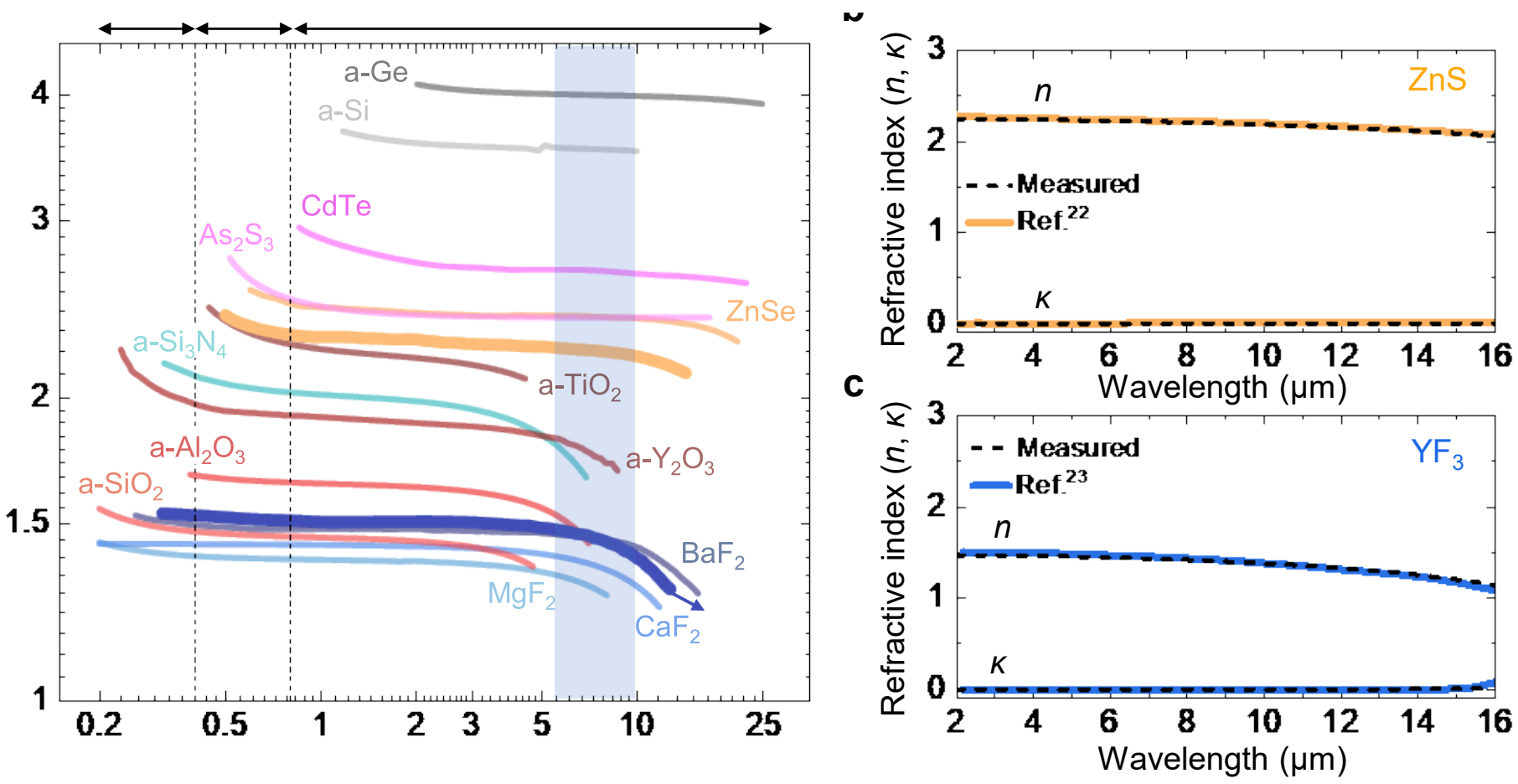

**Figure 3.** Potential low-loss optical materials for mid-IR ARCs. (a) Real part of the refractive index, $n$. Curves are plotted only in the wavelength regions where each material has an absorption coefficient $\alpha < 10$ cm$^{-1}$. ZnS and $YF_3$, which are used for the bilayer ARC in this work, are plotted with thicker lines. All data are from literature (ZnS[22], $YF_3$[23], ZnSe[22], a-Ge[24], a-Si[25], CdTe[26], $As_2S_3$[27], a-$Si_3N_4$[28], a-$TiO_2$[29], a-$Y_2O_3$[30], a-$Al_2O_3$[29], a-$SiO_2$[31], $BaF_2$[22], $CaF_2$[32], $MgF_2$[33], with a comprehensive summary in ref.[17]). (b, c) Real ($n$) and imaginary ($\kappa$) parts of the refractive index of e-beam evaporated ZnS and $YF_3$, measured using mid-IR variable-angle spectroscopic ellipsometry (IR-VASE) (dashed), compared to the literature data[22,23].

The planar bilayer design exhibits good tolerance to variations in film thickness. As shown in Fig. S4, varying the $YF_3$ thickness by ± 200 nm and the ZnS thickness by ± 150 nm leads to no significant changes in the reflectance. Additionally, by applying the planar bilayer ARCs to both the top and bottom surfaces of $BaTiS_3$ crystals, interference induced Fabry–Pérot fringes can be substantially mitigated within the broad design wavelength range [Fig. S5].

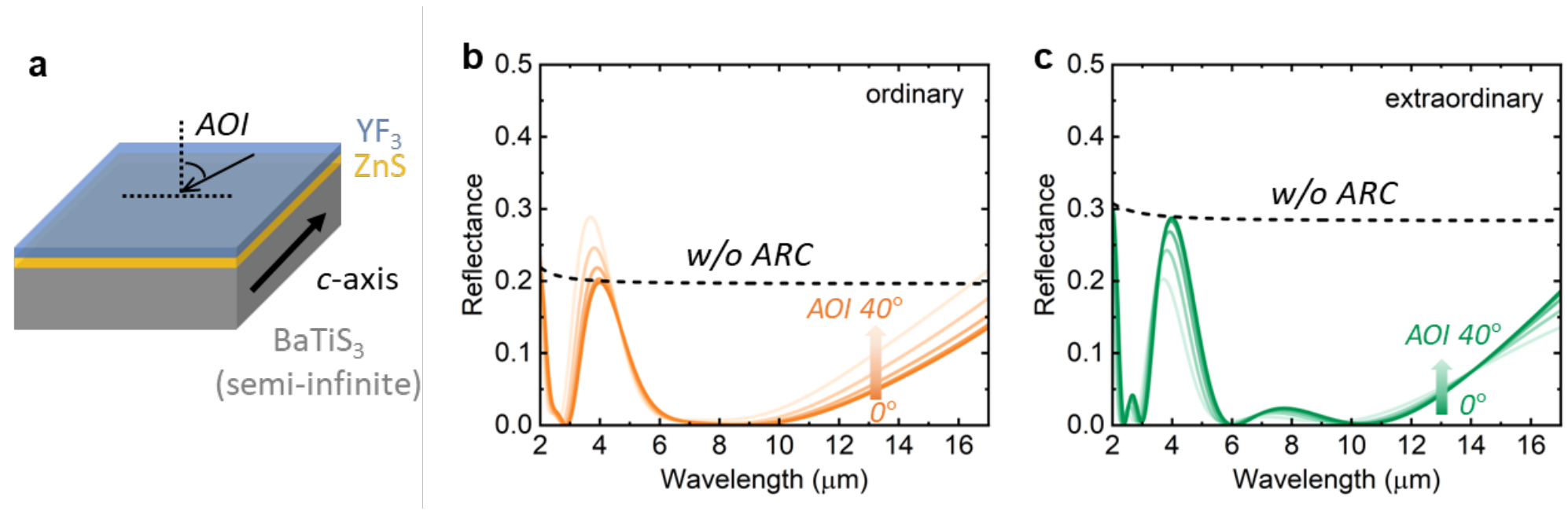

**Figure 4.** (a) Planar bilayer ARC on a $BaTiS_3$ substrate, with 0.85-μm-thick ZnS and 1.41-μm-thick $YF_3$. (b, c) Calculated reflectance at different angles of incidence (AOI) of the structure in (a), showing a small change for angles up to 40° in the range of 6–10 μm.

This simple planar bilayer ARC design also maintains robust performance at oblique incidence, remaining effective for angles exceeding 40°, as shown in Fig. 4 and Fig. S6. Such angular insensitivity is particularly important for practical optical systems, where finite numerical aperture, beam divergence, or off-axis illumination inevitably lead to a broad distribution of incidence angles.

Following the same design procedures, we explored planar bilayer ARC designs for $Sr_{9/8}TiS_3$, and we found that a $YF_3$/ZnSe combination provides the best suppression of reflectance for both polarizations [Figs. S7 and S8].

## 3. Experimental results

We realized the planar bilayer ARC using electron-beam evaporation of ZnS and $YF_3$ directly on top of a single crystal of $BaTiS_3$. The $BaTiS_3$ sample is 400 μm wide, ~1 mm long, and 350 μm thick, synthesized via the molten flux growth method[12], as shown in Fig. 5a.

The $BaTiS_3$ crystal was mounted on a glass carrier wafer, and loaded into a conventional e-beam evaporator (KVE-E2004, Korea Vacuum Tech.), with depositions performed at $10^{-6}$ Torr (more details in *Methods*). We evaporated 0.85 μm of ZnS directly on the $BaTiS_3$, followed by a 10-nm $Y_2O_3$ adhesion layer[34], and then the 1.46-μm-thick $YF_3$ film. Visually, the $YF_3$/ZnS layers appear to be continuous and mostly uniform [Fig. 5d]; the visible fringes indicate relatively small thickness differences across the sample compared to our target mid-IR wavelength. The complex refractive indices of $YF_3$ and ZnS were obtained by measuring and fitting ~100-nm-thick films of material that were electron-beam evaporated on a quartz wafer using mid-IR variable-angle spectroscopic ellipsometry (IR VASE, J. A. Woollam Co.), as shown in Figs. 3 (b, c). In our calculation, we also verified that the 10-nm $Y_2O_3$ adhesion layer has a negligible effect on the reflectance spectra.

We measured the reflectance of the same $BaTiS_3$ crystal with and without the ARC using an infrared microscope using an objective with numerical aperture (NA) = 0.4, integrated with a Fourier-transform infrared (FTIR) spectrometer. A wire-grid polarizer was used to control the polarization of the incident light, enabling measurement of reflectance along both the ordinary and extraordinary axes.

The measured reflectance spectra of the uncoated $BaTiS_3$ crystal are shown in Fig. 5b, with reflectance ~0.3 and ~0.2 for the extraordinary and ordinary directions, respectively. The measurement matches very well with calculations performed using previously published anisotropic complex refractive index data[11,13] [Fig. 5c].

Following deposition of the bilayer ARC, the measured reflectance was substantially suppressed over the 6 to 10 μm range, decreasing from 0.3 to ~0.04 for the extraordinary axis and from 0.2 to ~0.007 for the ordinary axis [Fig. 5e]. The measurements are in good agreement with the calculations [Figs. 5(c, f)]. The residual reflectance is slightly larger along the extraordinary axis, since the bilayer is optimized for the polarization-averaged reflectance and thus represents a compromise between the two axes. The higher-index extraordinary axis requires stronger suppression and retains a larger residual, and the same trend appears in our $Sr_{9/8}TiS_3$ designs [Fig. S7].

We note that while the sample in Fig. 5d is robust, surviving shipment and multiple measurements over many months, some other samples deposited in the same process resulted in delamination of the films. We attribute this delamination to thermal-expansion mismatch at two interfaces: between ZnS (coefficient of thermal expansion, CTE ~ $6.8 \times 10^{-6}$ $K^{-1}$)[35] and $YF_3$ (CTE ~ $16.9 \times 10^{-6}$ $K^{-1}$)[36], and between the coating stack and the $BaTiS_3$ substrate. Although we incorporated a $Y_2O_3$ adhesion layer (CTE ~ $8.0 \times 10^{-6}$ $K^{-1}$)[36], which is known to be helpful for preventing delamination[34], the thermal-expansion mismatch between the $BaTiS_3$ and the ARC is still present.

To better understand the mismatch with the $BaTiS_3$, we performed SC-XRD measurements of $BaTiS_3$, and found it to have a strongly anisotropic linear CTE value of ~$27.8 \times 10^{-6}$ $K^{-1}$ along the *a* and *b* axes and ~$13.2 \times 10^{-6}$ $K^{-1}$ along the *c*-axis [Figs. S9], resulting in a large biaxial strain in the deposited ZnS layer upon cooling from the deposition temperature, sometimes resulting in the delamination.

To mitigate the CTE mismatch, future ARC depositions may be performed at lower deposition temperatures, for example using plasma-enhanced processes[37]. Furthermore, optimizing the choice of adhesion layer can reduce interfacial stress[38].

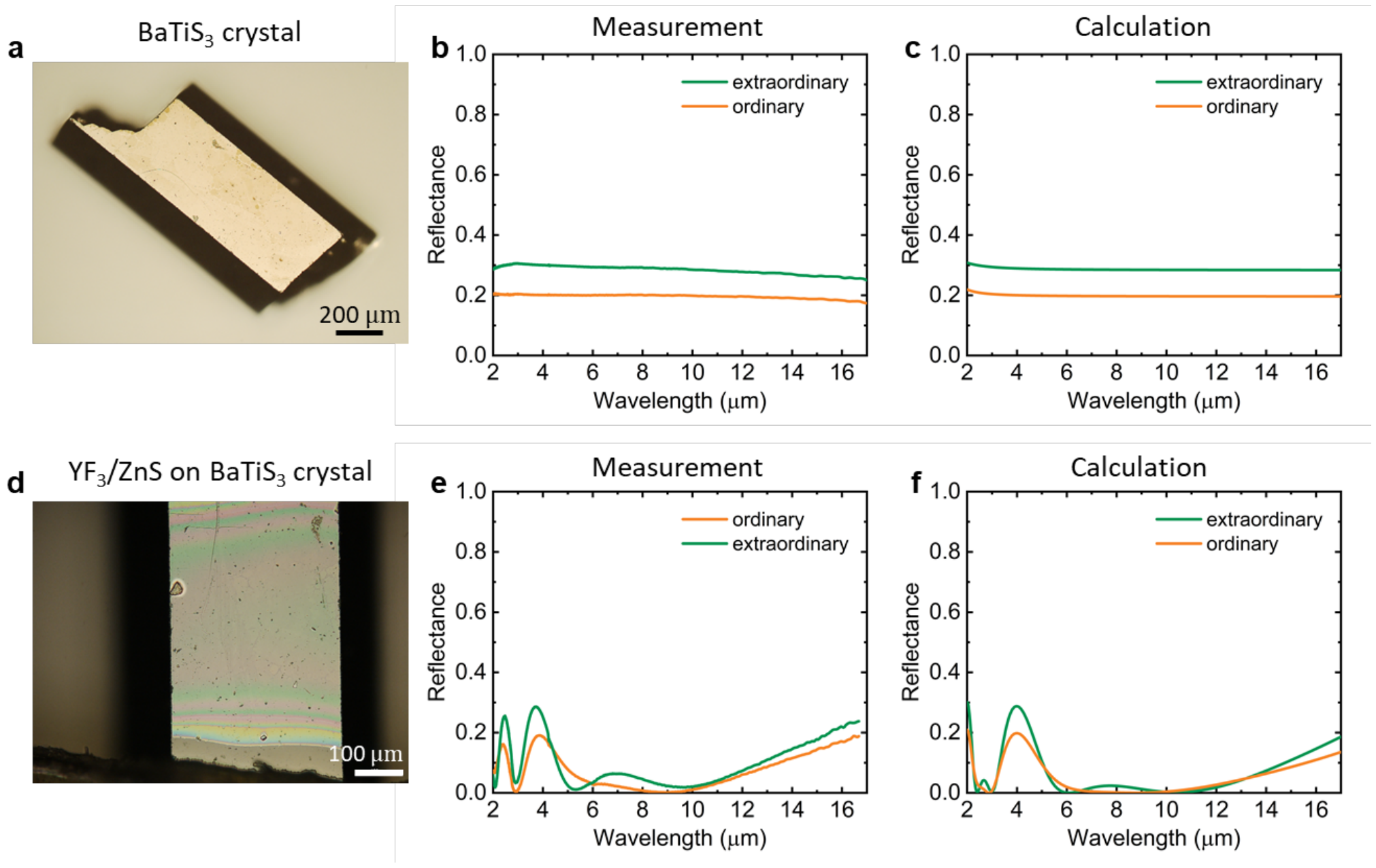

**Figure 5.** $BaTiS_3$ crystal (a–c) before and (d–f) after deposition of the ARC. (a, d) Optical images of the crystal surface. (b, e) Measured reflectance spectra. (c, f) Calculated reflectance spectra using the transfer-matrix method (TMM), using measured optical properties of $BaTiS_3$ and of $YF_3$ and ZnS from Figs. 3(b, c). Note that panel (c) is identical to Fig. 1c and is included here for comparison.

## 4. Conclusion

This study demonstrates the design and realization of anti-reflection coatings (ARCs) specifically tailored for highly anisotropic materials in the mid-infrared (mid-IR) region, focusing on quasi-one-dimensional $ABX_3$ crystals $BaTiS_3$ and $Sr_{9/8}TiS_3$ that have birefringence ($\Delta n$) on the order of unity. Owing to the long wavelengths in the mid-IR, we sought to minimize both the thickness and the number of ARC layers, and accordingly proposed two architectures: (i) a single-layer form-birefringent ARC and (ii) a planar bilayer ARC. We experimentally realized the planar bilayer ARC by electron-beam evaporation of yttrium fluoride ($YF_3$) and zinc sulfide (ZnS) onto $BaTiS_3$ and demonstrated substantial suppression of reflectance at this highly anisotropic interface for all polarizations.

The demonstration of ARCs for materials with giant anisotropy will enable the development of optical components such as ultra-thin mid-IR wave plates and polarizers, which would otherwise suffer from strong multiple-interface reflections.

## Methods

### *Single Crystal Growth*

Large single crystals of $BaTiS_3$ were synthesized using the molten flux method, with potassium iodide (KI) as the flux[14]. First, polycrystalline $BaTiS_3$ powders were prepared by solid-state reaction at 1040 °C, followed by annealing with excess sulfur (in a 2:1 $BaTiS_3$: S weight ratio) at 650 °C. The synthesized $BaTiS_3$ powders were then mixed with KI in a ~1:100 molar ratio, heated to 1040 °C, and then slowly cooled down to 700 °C (1 °C/h), after which the furnace was turned off. $BaTiS_3$ crystals were then recovered by washing off the solidified flux with DI water.

### *Thin film deposition*

To analyze the optical and material properties of ZnS (iTasco, 99.99%), $YF_3$ (iTasco, 99.99%), and $Y_2O_3$ (iTasco, 99.99%), ~100-nm-thick thin films of these materials were deposited on a quartz substrate (thickness 500 ± 30 μm, Hi-Solar Co., Ltd.). Before the deposition, the quartz substrates were sequentially cleaned using acetone, isopropyl alcohol, and distilled water with ultrasound, followed by a soft bake at 150 °C for 10 minutes to remove moisture and enhance adhesion.

Before depositing the ARCs, a rapid ICP-RIE was performed to clean the surface of the flux-grown $BaTiS_3$ samples. To enhance the adhesion of each of the thicker ZnS and $YF_3$ layers, a 10 nm layer of $Y_2O_3$ was deposited before the deposition of each material. The materials mentioned above were sequentially deposited on the samples using a conventional e-beam evaporator (KVE-E2004, Korea Vacuum Tech.) at $10^{-6}$ Torr. The deposition rates were set to 2 Å/sec for the ZnS film, 3 Å/sec for the $YF_3$ film, and 1 Å/sec for the $Y_2O_3$ film. During the deposition process, the substrate temperature was maintained at 150 °C, and a beam source sweep was utilized to ensure stable material deposition.

### *Optical measurement in infrared*

The complex refractive indices of $YF_3$ and ZnS were obtained by measuring and fitting ~100-nm-thick films of material using mid-IR variable-angle spectroscopic ellipsometry (IR VASE, J. A. Woollam Co.) over a spectral range of 2 to 20 μm at an angle of incidence of 50°, 60°, and 70°. Data analysis and refractive index extraction were performed using WVASE software (J. A. Woollam Co.).

Polarization-resolved infrared spectroscopy was carried out using a Fourier transfer infrared spectrometer (FTIR, Bruker Vertex 70) outfitted with an infrared microscope (Hyperion 2000). A 15× Cassegrain microscope objective with numerical aperture (NA) of 0.4 was used under normal incidence on the (100) face of the $BaTiS_3$ crystal. These measurements were performed with a Globar source, a potassium bromide (KBr) beam splitter, and a mercury cadmium telluride (MCT) detector. A wire-grid polarizer was used to control the polarization of the incident light. The samples were maintained at room temperature.

### *Measurement of the coefficients of thermal expansion (CTE) of $BaTiS_3$*

The CTE of $BaTiS_3$ was fitted by measuring the temperature-dependent lattice parameters of $BaTiS_3$ using single-crystal X-ray diffraction (SC-XRD) using a Rigaku XtaLAB Synergy-S diffractometer equipped with a HyPix-6000HE detector and Mo K$\alpha$ radiation (λ = 0.7107 Å). The $BaTiS_3$ crystal was mounted on a MiTeGen Dual Thickness MicroMount, and temperature control was achieved using an Oxford Cryosystems Cryostream 800. Measurements were performed over the temperature range of 255–325 K in 5 K steps during both heating and cooling cycles. A ramp rate of 6 K/min was used, with a 2-minute hold at each setpoint prior to data collection to allow equilibrium temperature. Data reduction and unit cell determination were performed using CrysAlisPro [Table 1]. The full data is shown in Fig. S9.

The linear CTE along each optical axis was extracted from a linear fit to the combined heating and cooling cycle data of the lattice parameter versus temperature, using $\alpha_L = \frac{1}{L}\frac{\Delta L}{\Delta T}$ where L is the lattice parameter and $\Delta L/\Delta T$ is the slope of the linear fit[39]. The volumetric CTE was similarly determined from a linear fit to the combined heating and cooling cycle data of the unit cell volume versus temperature using $\alpha_V = \frac{1}{V}\frac{\Delta V}{\Delta T}$, where $\Delta V/\Delta T$ is the slope of the linear fit to the unit cell volume versus temperature data. $\alpha_V$ is equivalent to the sum of the linear CTEs along all three optical axes, i.e., $\alpha_V = \alpha_a + \alpha_b + \alpha_c = \sim 68.8 \times 10^{-6}\ K^{-1}$.

**Table 1:** Data reduction statistics for measurement taken at 255 K during cooling.

| Diffractometer | Rigaku XtaLAB Synergy-S | |
|---|---|---|
| Source | PhotonJet (Mo) X-ray Source | |
| Wavelength | 0.71073 Å | |
| Temperature | 255 K | |
| Laue group | $6/mmm$ | |
| **Cell dimensions** | | |
| $a$, $b$, $c$ (Å) | 11.6832(13), 11.6832(13), 5.7976(7) | |
| α, β, γ (°) | 90, 90, 120 | |
| Volume (Å$^3$) | 685.34(13) | |
| **Integration and Scaling** | | |
| Resolution (Å) | 2θ ~ Inf. to 0.70 | 2θ ~ Inf. to 0.80 |
| Completeness (%) | 85.2 | 99.7 |
| Redundancy | 7.3 | 8.1 |
| Mean $F^2$ / σ($F^2$) | 24.47 | 25.87 |
| $R_{int}$ | 0.024 | 0.023 |
| $R_{\sigma B}$ | 0.014 | 0.013 |
| **Statistics of Reflections** | | |
| $h_{min}$, $h_{max}$ | -14, 13 | |
| $k_{min}$, $k_{max}$ | -12, 16 | |
| $l_{min}$, $l_{max}$ | -7, 6 | |

## Funding

The work at UW-Madison was supported by Office of Naval Research (ONR), award no. N00014-20-1-2297 during the earlier stages of the research, by the National Science Foundation (NSF), award number DMR-2522863, during the later stages.

The work at Kyung Hee University was supported by National Research Foundation of Korea through the Basic Science Research Program (RS-2026-25469967).

The crystal growth and characterization, in part, was supported by the Army Research Office (ARO) under award numbers W911NF-21-1-0327 (MURI) and W911NF-24-1-0164 and US National Science Foundation grants with award numbers DMR-2122071 and DMR-2522862. The crystal growth and characterization tools were supported in part by an ONR grant with award number N00014-23-1-2818.

## Acknowledgements

H. Mei, J.-W. Cho, and M. A. Kats gratefully acknowledge use of facilities and instrumentation in the UW-Madison Wisconsin Center for Nanoscale Technology (wcnt.wisc.edu), which is partially supported by the Wisconsin Materials Research Science and Engineering Center (NSF DMR-2309000) and the University of Wisconsin-Madison. NSF grant CHE-2018740 provided funds to acquire the Rigaku XtaLAB Synergy-S diffractometer used for single-crystal XRD studies.

## Disclosures

The authors declare no conflicts of interest.

## Data Availability

The data that support the findings of this study are openly available in Zenodo.

## Supplemental document

See Supplementary Information for supporting content.

# Supplementary Information

# Anti-reflection coatings for highly anisotropic materials in the mid infrared

**Hongyan Mei,[1,†] Jin-Woo Cho,[1,†] Jae-Seon Yu[2], Huandong Chen[3], Shantanu Singh[3,4], Boyang Zhao[3], Jayakanth Ravichandran[3,4], Sun-Kyung Kim[2], Mikhail A. Kats[1,*]**

*[1] Department of Electrical and Computer Engineering, University of Wisconsin-Madison, Madison, WI 53706, USA*
*[2] Department of Applied Physics, Kyung Hee University, Gyeonggi-do 17104, Republic of Korea*
*[3] Mork Family Department of Chemical Engineering and Materials Science, University of Southern California, Los Angeles, CA 90089, USA*
*[4] Core Center for Excellence in Nano Imaging, University of Southern California, Los Angeles, CA 90089, USA*
*[†] These authors contributed equally.*
*[*] mkats@wisc.edu*

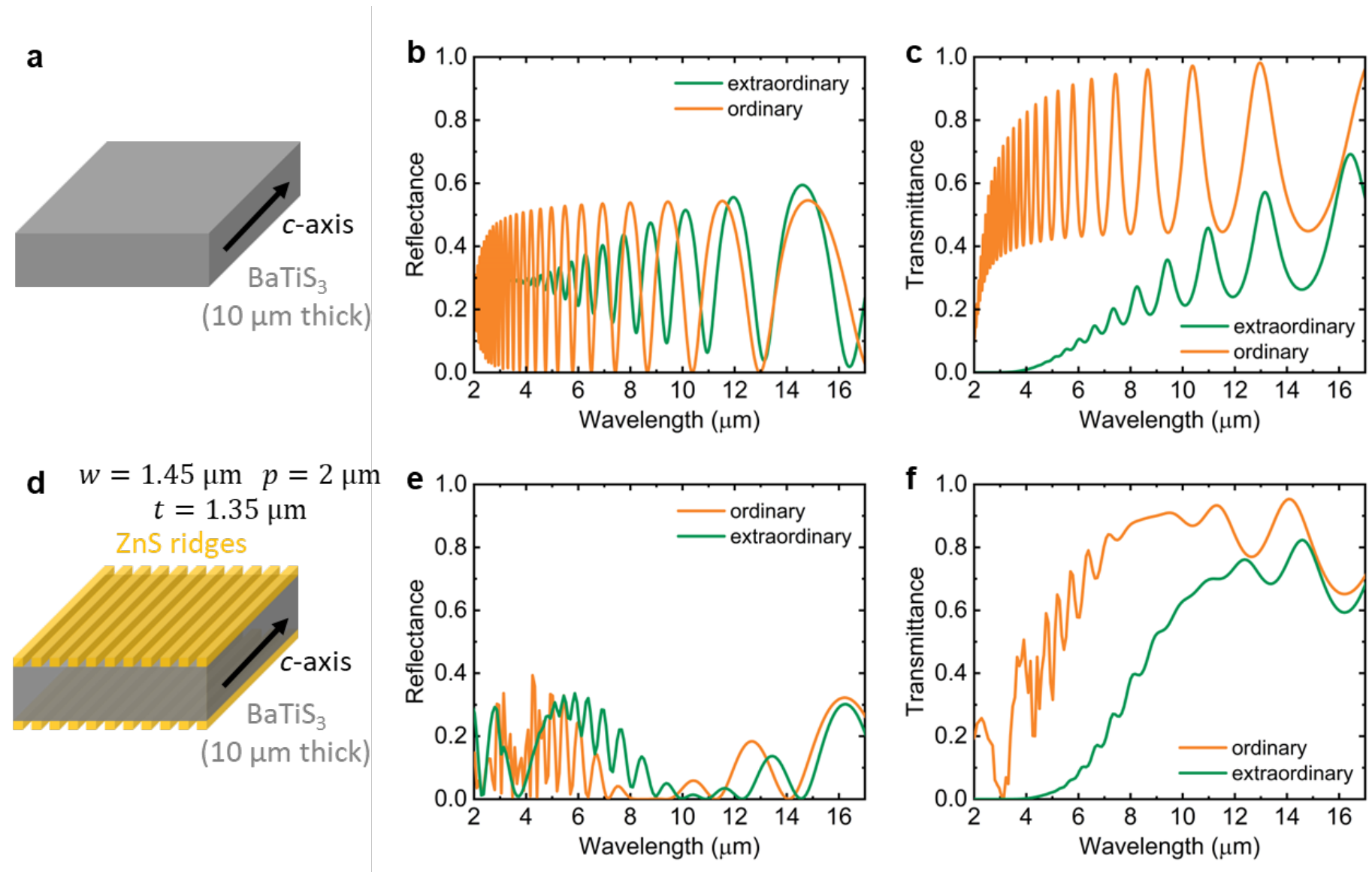

**Figure S1.** (a) Schematic of a 10-µm-thick $BaTiS_3$ crystal. (b, c) Calculated normal incidence reflectance and transmittance of the structure in (a), showing pronounced Fabry–Pérot fringes. (d) ZnS ridges (1.45 µm width, 2 µm period, 1.35 µm thickness) on both top and bottom surfaces of a 10-µm thick $BaTiS_3$ crystal, with the ridge direction along the *c*-axis. (e, f) Calculated normal incidence reflectance and transmittance of the structure in (d), showing suppression of reflectance, increase of the transmission, and mitigation of Fabry–Pérot fringes at the designed wavelength $\lambda = 10$ µm for both polarizations.

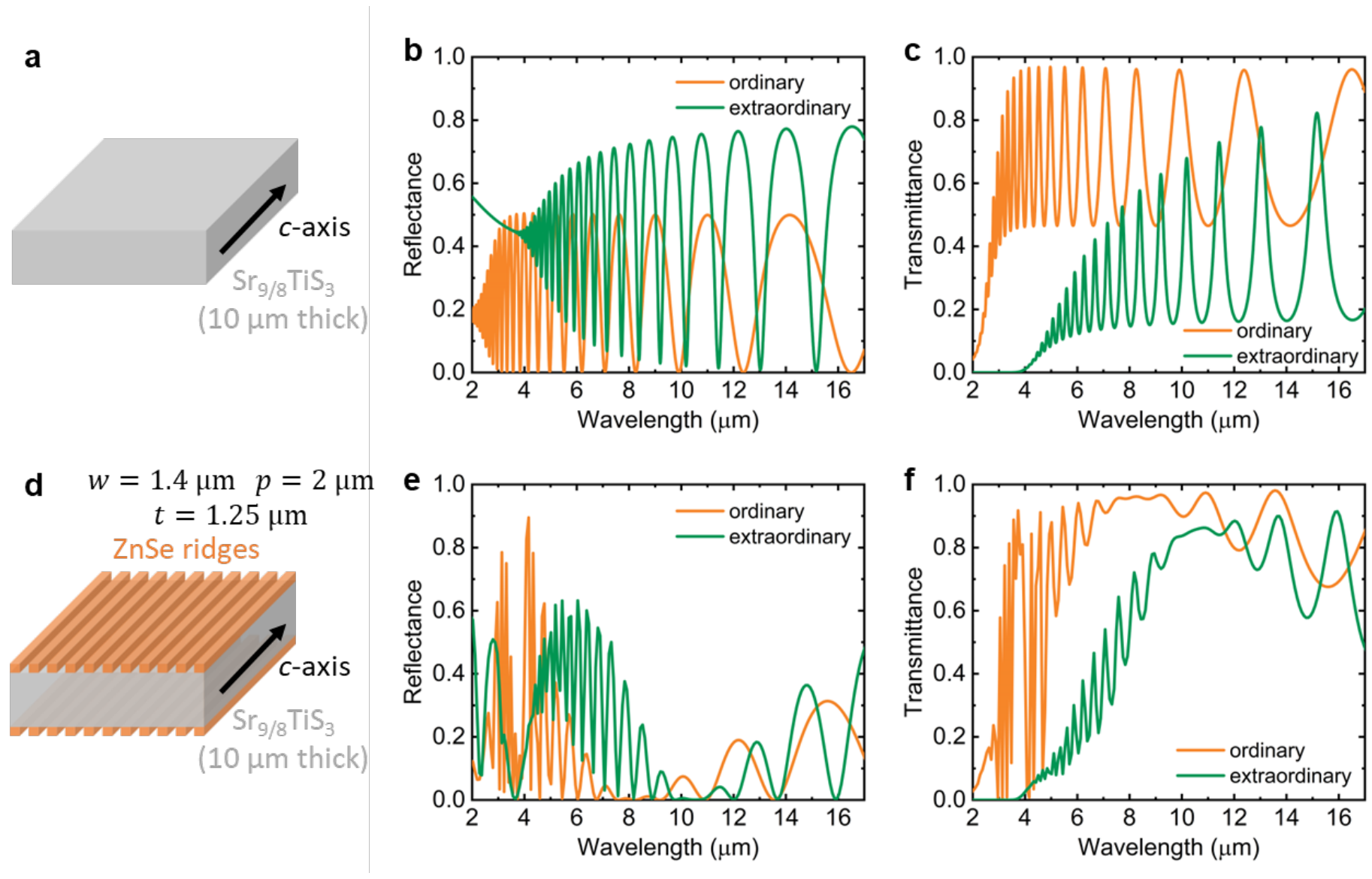

**Figure S2.** (a) Schematic of a 10-μm-thick $Sr_{9/8}TiS_3$ crystal. (b, c) Calculated normal incidence reflectance and transmittance of the structure in (a), showing pronounced Fabry–Pérot fringes. (d) ZnSe ridges (1.4 μm width, 2 μm period, 1.25 μm thickness) on both top and bottom surfaces of a 10-μm-thick $Sr_{9/8}TiS_3$ crystal, with the ridge direction along the *c*-axis. (e, f) Calculated normal incidence reflectance and transmittance of the structure in (d), showing suppression of reflectance, increase of the transmission, and mitigation of Fabry–Pérot fringes at the designed wavelength λ = 10 μm for both polarizations.

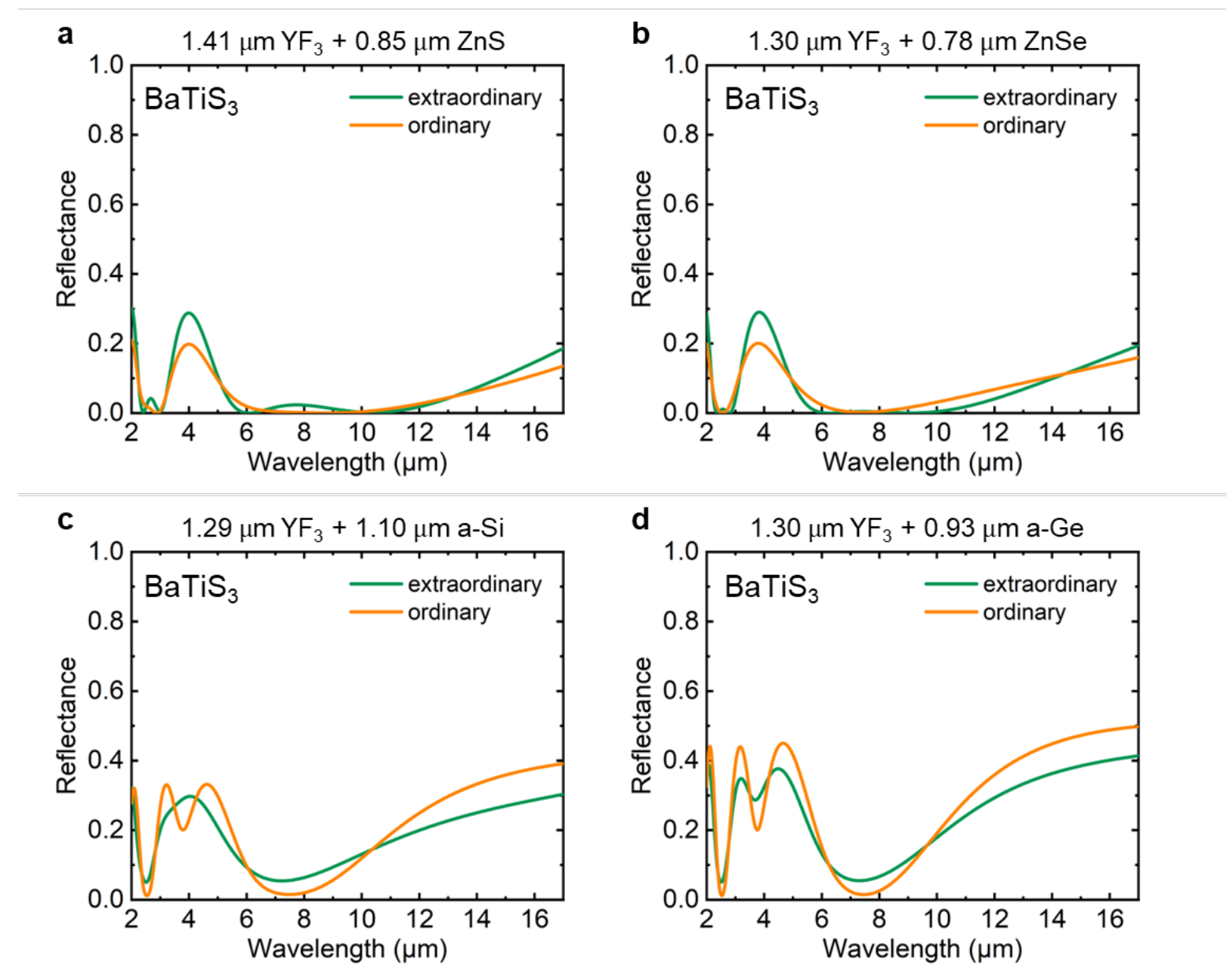

**Figure S3.** Calculated normal incidence reflectance of planar bilayer ARCs on a $BaTiS_3$ substrate, with a combination of the low-index material $YF_3$ and high-index materials (a) ZnS[22], (b) ZnSe[22], (c) a-Si[25], and (d) a-Ge[24]. The thicknesses of the layers are optimized to minimize the normal-incidence reflectance averaged over the two polarizations, targeting λ = 6–10 μm.

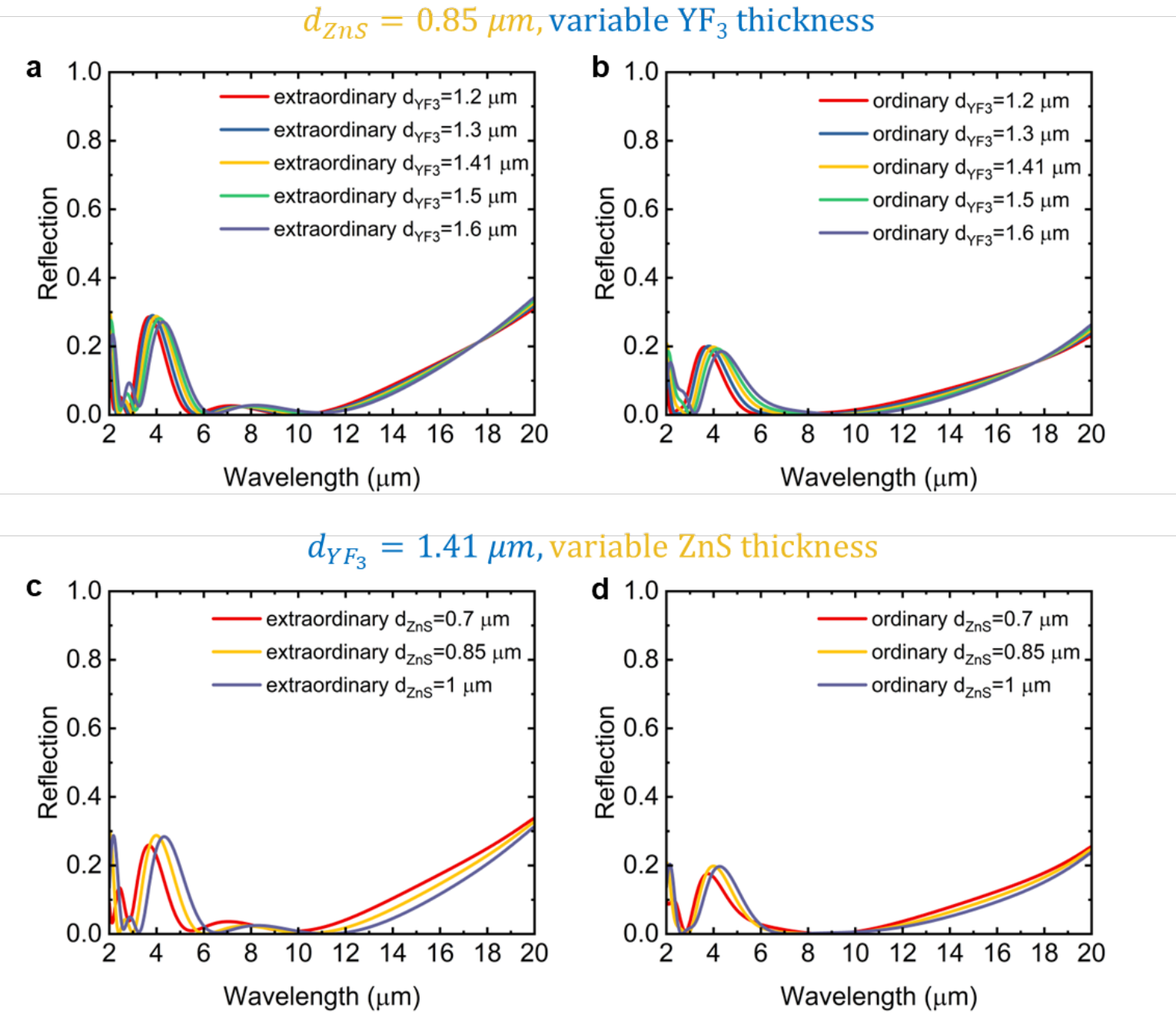

**Figure S4.** Thickness tolerance test for the double-layer $YF_3$/ZnS ARC for $BaTiS_3$. (a, b) Calculated normal incidence reflectance when varying the thickness of $YF_3$ and (c, d) varying the thickness of ZnS, showing small changes in reflectance.

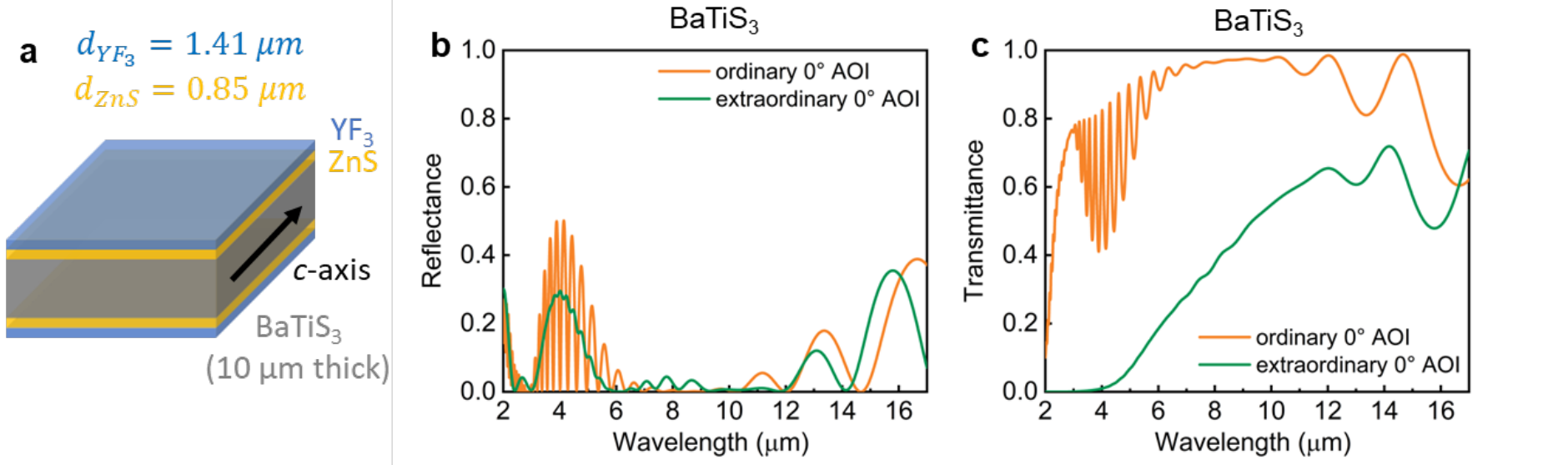

**Figure S5.** (a) Double-layer ARC (1.41-μm-thick $YF_3$ + 0.85-μm-thick ZnS) on both top and bottom surfaces of a 10-μm-thick $BaTiS_3$ crystal. (b, c) Calculated normal incidence reflectance and transmittance of the structure in (a), showing suppression of reflectance, increase of the transmission, and mitigation of Fabry–Pérot fringes at the designed wavelength range of 6–10 μm for both polarizations. Note that the transmittance and reflectance do not add up to 1 due to optical absorption in $BaTiS_3$.

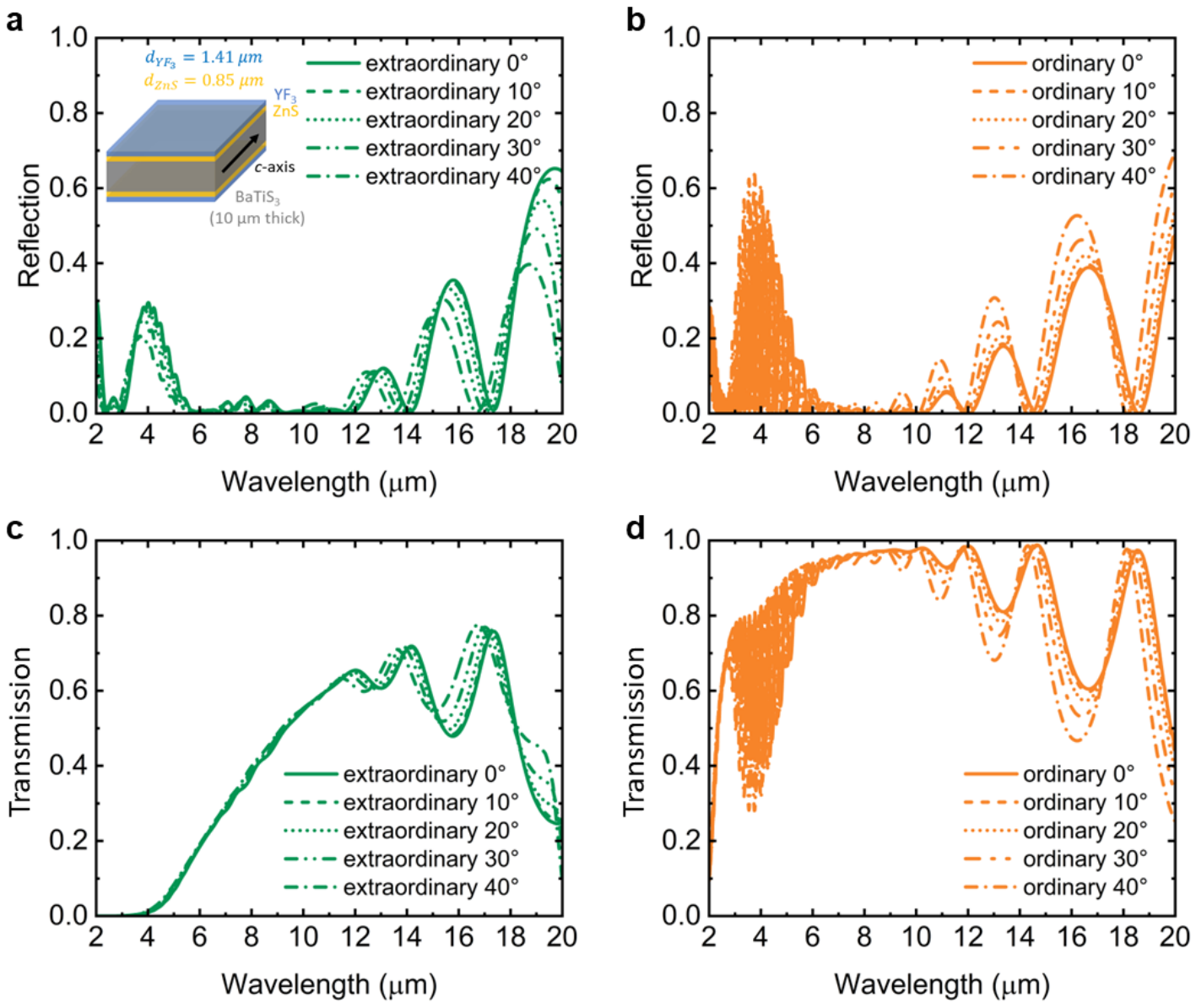

**Figure S6.** (a, b) Calculated reflectance and (c, d) transmittance at different angles of incidence for the structure shown in the inset of (a), showing only small changes for angles up to 40° in the wavelengths of 6–10 μm. The inset in (a) shows a double-layer ARC (1.41-μm-thick $YF_3$ + 0.85-μm-thick ZnS) on both top and bottom surfaces of a 10-μm-thick $BaTiS_3$ crystal. Note that the transmittance and reflectance do not add up to 1 due to optical absorption in $BaTiS_3$.

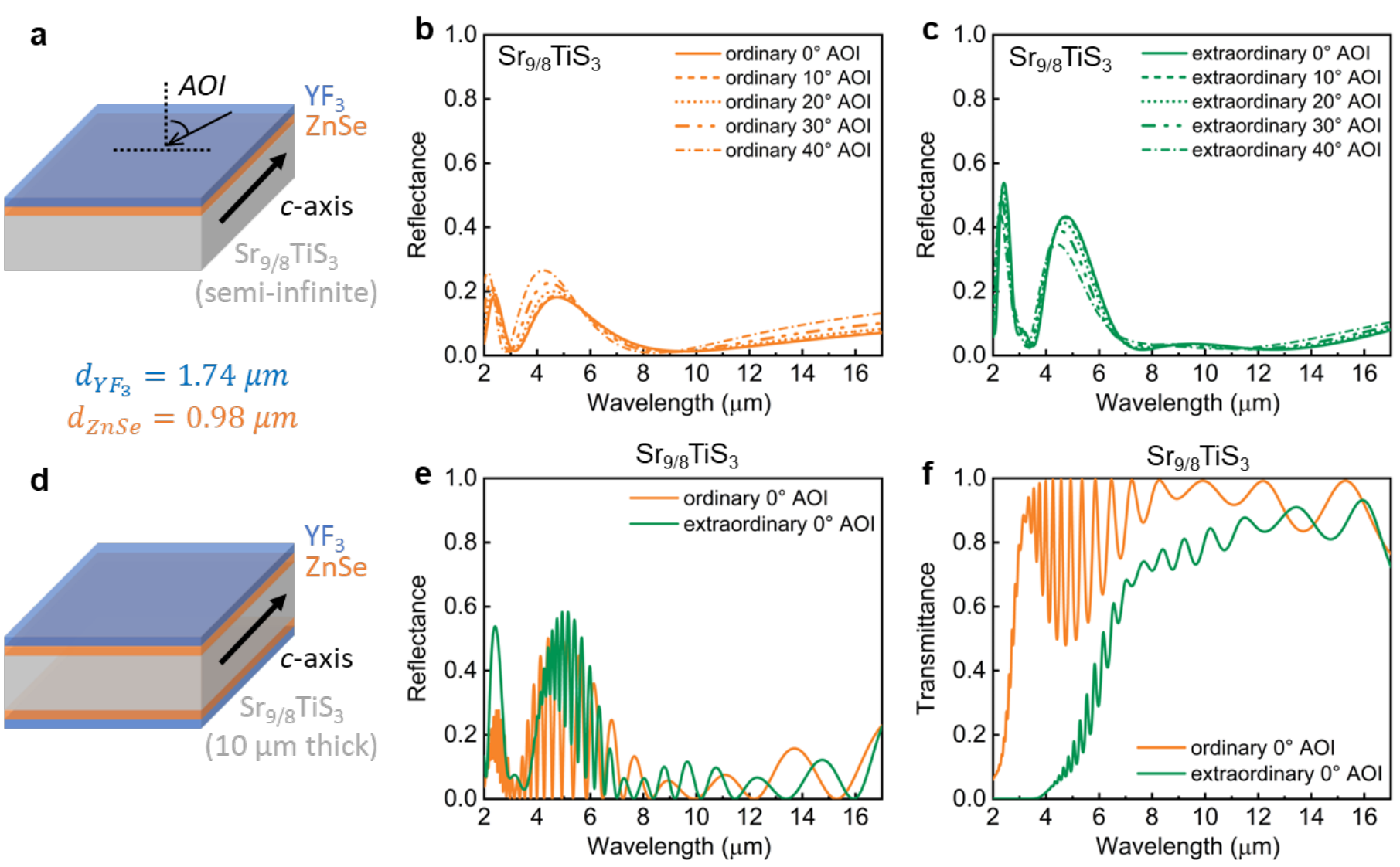

**Figure S7.** (a) Planar bilayer ARC on a $Sr_{9/8}TiS_3$ substrate, with high-index material ZnSe and low-index material $YF_3$. (b, c) Calculated reflectance at different angles of incidence of the structure in (a), showing a small change for angles up to 40° in the range of 6–10 μm. (d) Planar bilayer ARC (1.74-μm-thick $YF_3$ + 0.98-μm-thick ZnSe) on both top and bottom surfaces of a 10-μm thick $Sr_{9/8}TiS_3$ crystal. (e, f) Calculated reflectance and transmittance of the structure in (a), showing suppression of reflectance, increase of the transmission, and mitigation of Fabry–Pérot fringes at the designed wavelength range $\lambda$ = 6–10 μm for both polarizations. Note that the transmittance and reflectance do not add up to 1 due to optical absorption in $Sr_{9/8}TiS_3$.

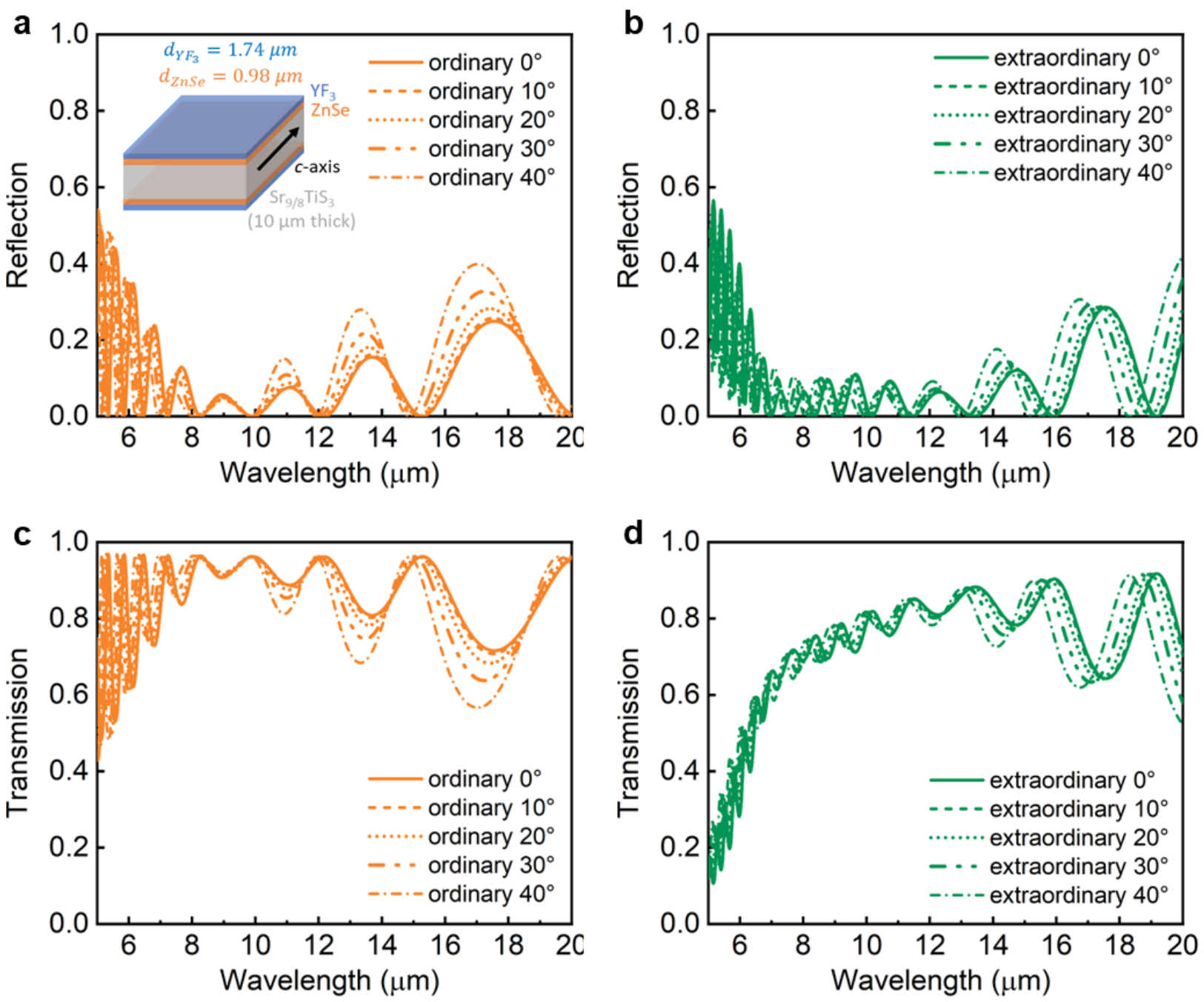

**Figure S8.** (a, b) Calculated reflectance and (c, d) transmittance at different angles of incidence for the structure shown in the inset of (a), showing only small changes for angles up to 40° in the range of 6–10 μm. The inset in (a) shows a double-layer ARC (1.74-μm-thick $YF_3$ + 0.98-μm-thick ZnSe) on both top and bottom surfaces of a 10-μm-thick $Sr_{9/8}TiS_3$ crystal. Note that the transmittance and reflectance do not add up to 1 due to optical absorption in $Sr_{9/8}TiS_3$

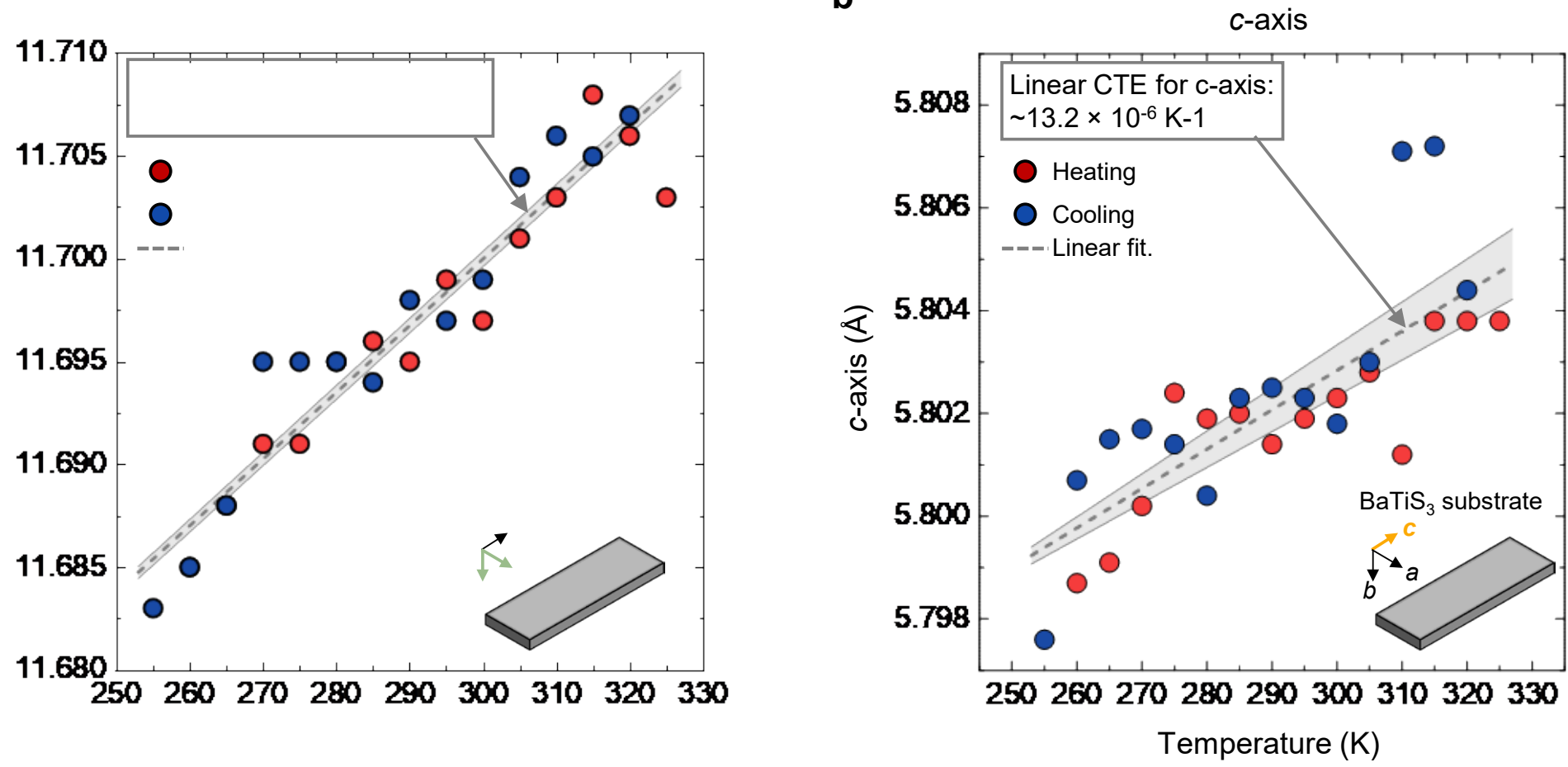

**Figure S9.** Temperature dependence of the $BaTiS_3$ lattice parameters determined from SC-XRD measurements over a 255–325 K heating and cooling cycles. (a,b) Temperature-dependent lattice parameters of (a) *a* and *b* axes (which are equivalent) and (b) the c-axis, with linear fits. The gray color band corresponds to the standard deviation between the heating and the cooling cycles. (Insets) Schematics of the $BaTiS_3$ crystal and the coordinate axes.